\documentclass[preprint]{revtex4}
\usepackage{graphicx}
\usepackage{amsmath}
\usepackage{amssymb}

\begin{document}

\title{Nonequilibrium interactions between two quantum circuits}

\author{V.S.~Khrapai$^{1,2}$, S.~Ludwig$^1$, J.P.~Kotthaus$^1$ and
H.P.~Tranitz$^3$, W.~Wegscheider$^3$}

%
%

\affiliation{$^1$ Center for NanoScience and Department
f$\ddot{\text{u}}$r Physik,
Ludwig-Maximilians-Universit$\ddot{\text{a}}$t,
Geschwister-Scholl-Platz 1, D-80539 M$\ddot{\text{u}}$nchen,
Germany; $^2$ Institute of Solid State Physics RAS, Chernogolovka,
142432, Russian Federation; $^3$ Institut f$\ddot{\text{u}}$r
Experimentelle und Angewandte Physik, Universit$\ddot{\text{a}}$t
Regensburg, D-93040 Regensburg, Germany}

\begin{abstract}
We briefly overview our recent results on nonequilibrium
interactions between neighbouring electrically isolated
nanostructures. One of the nanostructures is represented by an
externally biased quantum point contact (drive-QPC), which is used
to supply energy quanta to the second nanostructure (detector).
Absorption of these nonequilibrium quanta of energy generates a
dc-current in the detector, or changes its differential
conductance. We present results for a double quantum dot, a single
quantum dot or a second QPC placed in the detector circuit. In all
three cases a detection of quanta with energies up to $\sim$1~meV
is possible for bias voltages across the drive-QPC in the mV
range. The results are qualitatively consistent with an energy
transfer mechanism based on nonequilibrium acoustic phonons.
\end{abstract}

 \maketitle

\section{Introduction}

Present GaAs fabrication techniques enable one to create a pair of
nanostructures connected to separate two-dimensional electron gas
(2DEG) leads and placed just about 100~nm apart. Out of
thermodynamic equilibrium a net transfer of energy between such
two quantum circuits can occur. This can happen both directly via
a Coulomb interaction between the electrons of the two circuits,
and indirectly, via emission/absorption of energy quanta into/from
their common environment. In the last case, exchange with the
quanta of the electromagnetic field (photons) as well as those of
the crystal lattice vibrations (phonons) is possible thanks to
electromagnetic and electron-phonon interactions. Recent
experiments~\cite{Onac,DQDratchet,gustavsson} gave no definite
answer on what determines the dominant interaction mechanism in
similar devices. It is important to know this, e.g., for
application of coupled nanostructures in quantum measurements.

Regardless the type of interaction, the change of the energy and
momentum of an electron satisfies the conservation laws, which can
impose constraints for the respective energy transfer mechanism.
These constraints are most crucial for freely moving electrons.
For Coulomb interaction, e.g., the conservation of momentum
determines a positive sign of the Coulomb drag between clean
one-dimensional (1D) quantum wires~\cite{zverev} and parallel
2DEGs in bilayer systems~\cite{gramila}. Emission/absorption of an
energy quantum from the environment by a 2DEG electron is possible
provided the velocity $v$ of the corresponding particle (a photon
or an acoustic phonon here) is smaller than the electron's Fermi
velocity $v < v_F$~\cite{gantmakher}. This condition is only
fulfilled for acoustic phonons thanks to a small sound velocity
($v_s\ll v_F$ in typical 2DEGs). Conservation laws allow the
interaction with acoustic phonons of in-plane momenta as high as
$2k_F$ and corresponding energies up to $2\hbar k_Fv_s\sim$~1~meV,
where $k_F$ and $\hbar$ are, respectively, the Fermi momentum in
the 2DEG, and the Plank's constant. No strict constraints exist
for confined electrons because of a lack of momentum conservation.
Hence, the electrons in a quantum dot (QD) can interact both with
microwave photons~\cite{PAT,aguado} and acoustic
phonons~\cite{fedichkin}.

\begin{figure}[p]
\begin{center}
\includegraphics[clip,width=0.9\linewidth]{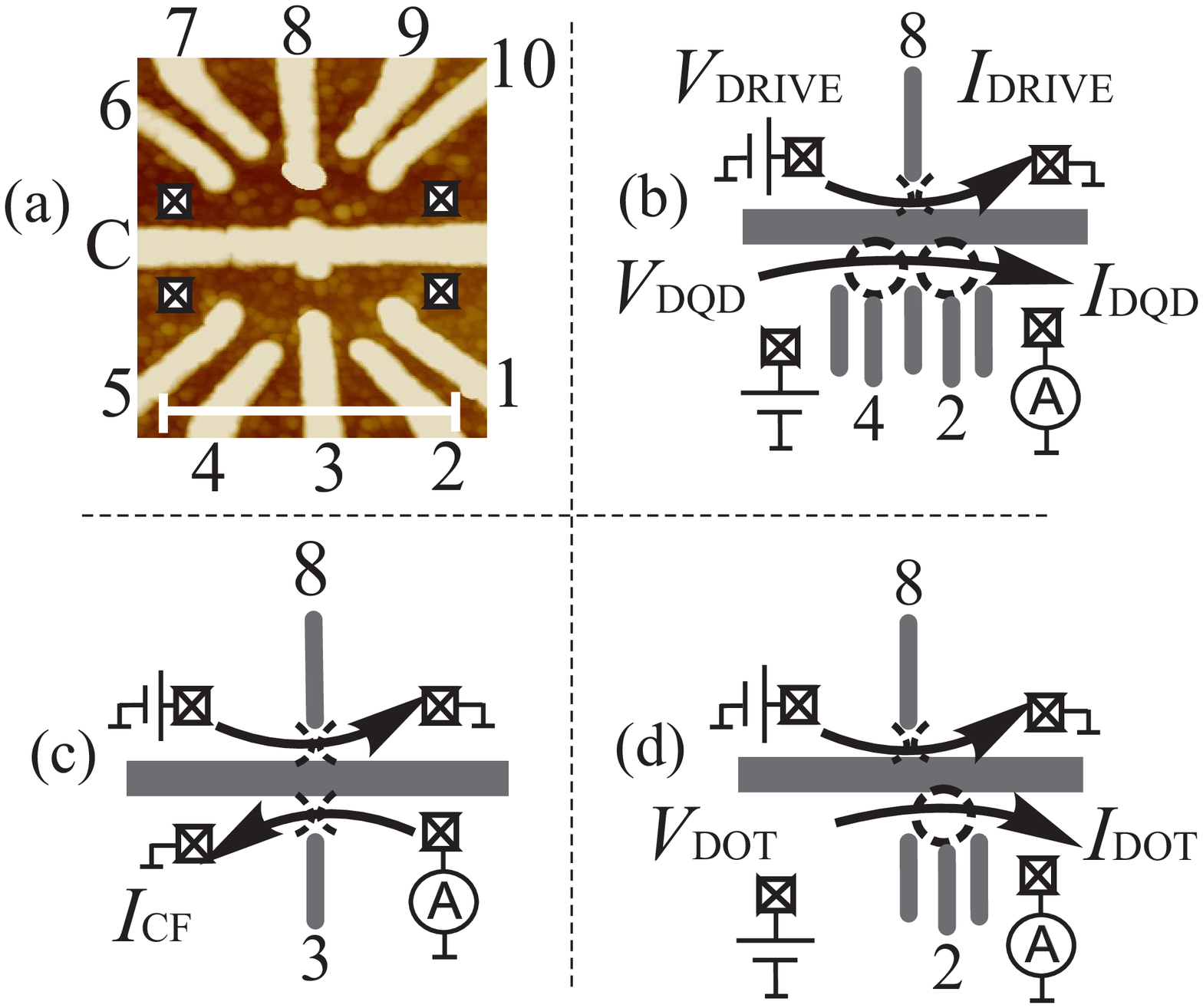}
\caption{(a): AFM micrograph of the nanostructure. Metal gates on
the surface of the heterostructure are shown in bright tone.
Crossed squares mark contacted 2DEG regions. The scale bar equals
1~$\mu$m. (b),(c),(d): sketches of the measurement configuration
for three different realizations of the detector nanostructure.
Gates used in each case are shown in dark.} \label{fig1}
\end{center}
\end{figure}
In this paper we overview a set of experiments (partly reported in
Refs.~\cite{DQDratchet,counterflowpaper,ep2dsprocs}) on
nonequilibrium interactions between neighbouring electrically
isolated nanostructures laterally defined within the 2DEG beneath
the surface of a GaAs/AlGaAs heterostructure. Compared to previous
work, we present new experimental data and extend a microscopic
discussion of the observations. The AFM micrograph of the sample
is shown in Fig.~\ref{fig1}a. The negatively biased central gate C
depletes the underlying 2DEG  and divides the sample into two
coplanar nanostructures, defined and controlled by voltages on
gates 1-10, with four separately contacted 2DEG leads (marked by
crossed squares in fig.~\ref{fig1}a). One of the nanostructures is
an externally biased quantum point contact (drive-QPC) and is used
to supply energy to the second nanostructure (detector).
Absorption of energy results in generation of a dc current in the
detector circuit or changes it's differential conductance, which
can be measured in the experiment. The energy spectrum of the
excitation as well as its spatial asymmetry are studied by using a
double quantum dot (DQD) (fig.~\ref{fig1}b), a second QPC
(fig.~\ref{fig1}c) or a single QD (fig.~\ref{fig1}d) as the
detector. In all three cases the detection of quanta with energies
up to $\sim$1~meV occurs for bias voltages across the drive-QPC
($V_\text{DRIVE}$) in the mV range. As shown below, our
observations demonstrate that the drive-QPC provides a strong
spatially asymmetric excitation to the electrons of the 2DEG leads
of the detector. This strongly suggests that the dominant energy
transfer mechanism between the two quantum circuits in our
experiment is based on emission/absorption of nonequilibrium
acoustic phonons happening in the 2DEG leads of the drive/detector
nanostructure. This mechanism has to be considered in experiments
on coupled quantum circuits, at least in the regime of high
external bias.

The paper is organized as follows. The details of the experiment
are described in section~\ref{exp}. In the subsequent sections the
results for three detector realizations are presented. In
section~\ref{DQD} we describe the experiment with the
DQD-detector, which provides a quantitative measure for the
drive-QPC mediated excitation bandwidth. Observation of a
so-called counterflow effect~\cite{counterflowpaper} with the
detector-QPC is described in section~\ref{counterflow}. A
qualitative analogy as well as a strong quantitative difference of
the results to thermopower experiments in single
QPCs~\cite{molenkamp,dzurak} are given in this section. Excitation
of discrete energy levels in the QD-detector mediated by the
drive-QPC is reported in section~\ref{QD}. The discussion of the
observations in terms of a phonon-mediated energy transfer
mechanism between the two circuits is given in the last
section~\ref{conclusions}.

\section{Experimental details}\label{exp}

All the measurements presented below were performed on a
GaAs/AlGaAs heterostructure, containing a 2DEG 90~nm below the
surface, with a carrier density of $2.8\times10^{11}$~cm$^{-2}$
and a low-temperature mobility of $1.4\times10^6$~cm$^2/$Vs. The
metallic gate layout of fig.~\ref{fig1}a was designed by means of
e-beam lithography. The sample was immersed in the mixing chamber
of a dilution refrigerator with a base temperature of 25~mK and
cooled down to an electron temperature below 150~mK. Dc or low
frequency (21~Hz) ac current measurements in the drive and
detector circuits were performed by use of two current-voltage
converters with variable gain from $10^{6}$ to $10^{9}$ V/A
followed by a digital voltmeter or a lock-in amplifier,
respectively. The lock-in technique was particularly useful for
low-impedance counterflow measurements (see
section~\ref{counterflow}), where the dc signal to noise ratio was
poor. In some cases, a differential signal was obtained by
numerically deriving the dc current data (section~\ref{QD}) or the
dc data were obtained via numerical integration of the ac signal
(section~\ref{counterflow}). We have carefully checked that these
procedures are equivalent in the regime of the nearly pinched-off
detector-QPC. Careful check for absence of the leakage between the
two circuits, measurements with interchanged signal and ground
ohmic contacts, interchanged drive and detector nanostructures, as
well as simultaneous dc and ac measurements, were performed to
ensure the small signals measured are free from spurious effects.

\section{Double-dot quantum ratchet}\label{DQD}

In this section we describe the experiment with a DQD in the
detector circuit~\cite{DQDratchet}. A sketch of the measurement is
shown in fig.~\ref{fig1}b. Two serially connected QDs with weak
interdot coupling ($t\sim0.1~\mu$eV) and strong dot-lead coupling
($\Gamma\approx40~\mu$eV) are formed on one side of the gate C by
negatively biased gates 1-5. Typical values of charging energy,
single-particle level spacing and interdot Coulomb energy are,
respectively 1.5~meV, 100~$\mu$eV and $100-200~\mu$eV. The charge
configuration of the DQD is controlled by voltages $V_2,~V_4$
applied to gates 2 and 4, which predominantly couple to the
electrochemical potentials of the right and left QD, respectively.
A small bias voltage across the DQD $V_\text{DQD}=-20~\mu$V is
applied throughout the experiment.

\begin{figure}[p]
\begin{center}
\includegraphics[clip,width=0.4\linewidth]{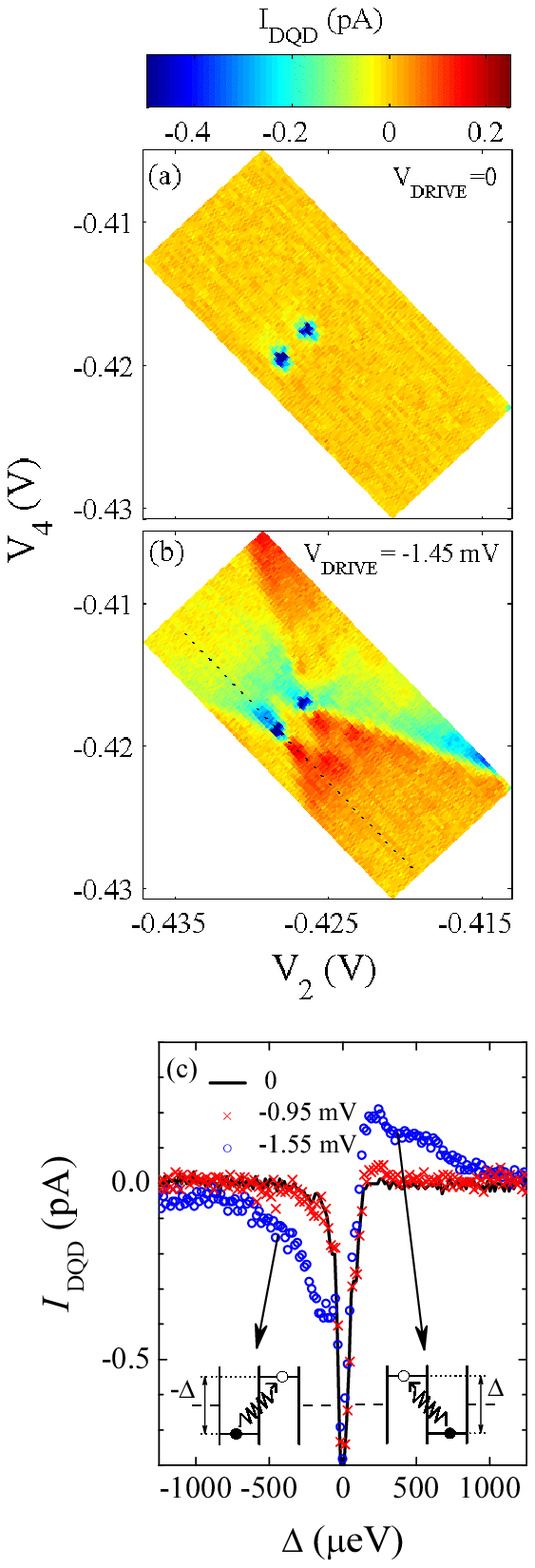}
\caption{(a),(b): Colour-scale plot of $I_\text{DQD}$ for two
corresponding values of $V_\text{DRIVE}$. The top colour-bar is
the same for both panels. (c): $I_\text{DQD}$ vs the
electrochemical potential difference between the two dots, taken
along the dashed trace in (b) for three values of
$V_\text{DRIVE}$. The left/right inset schematically shows the
inelastic interdot tunnelling processes responsible for
negative/positive ratchet contribution to $I_\text{DQD}$.
Figure~\ref{fig2}c is reproduced with permission from
Ref.~\cite{ep2dsprocs} with minor changes to axes scales.}
\label{fig2}
\end{center}
\end{figure}
In the absence of current in the drive-circuit
($V_\text{DRIVE}=0$), $I_\text{DQD}$ is mainly suppressed because
of the Coulomb blockade (fig.~\ref{fig2}a). The only exception is
a pair of sharp resonances in the $[V_2,V_4]$ plane (so called
stability diagram), where the electrochemical potentials of both
dots $\mu_R,~\mu_L$ and 2DEG leads are aligned~\cite{vanderwiel}.

The situation changes drastically at finite bias across the
drive-QPC, tuned halfway between the pinch-off and first
conductance plateau ($g_\text{DRIVE}\equiv
dI_\text{DRIVE}/dV_\text{DRIVE}\approx0.5~G_0$, where $G_0=2e^2/h$
is the conductance quantum). In figure~\ref{fig2}b $I_\text{DQD}$
is plotted throughout the same region of the stability diagram for
$V_\text{DRIVE}=-1.45$~mV. In contrast to fig.~\ref{fig2}a, now a
non-zero current flows across the DQD in the regions of stable
ground state charge configurations. The sign of the DQD current
depends on the position in the stability diagram relative to the
resonances. The current is negative on the left and above the
resonances and positive on the right and below them
($I_\text{DQD}>0$ corresponds to electrons moving to the left-hand
side in the lower circuit of fig.~\ref{fig1}b). $I_\text{DQD}$
changes abruptly at the boundaries of stable ground state
configurations, making them visible in fig.~\ref{fig2}b (nearly
horizontal and vertical lines originating from resonances, see
Ref.~\cite{DQDratchet} for details). Note that such a behaviour is
observed around many pairs of resonances in the stability
diagram~\cite{ep2dsprocs}.

All the main features of fig.~\ref{fig2}b can be explained by
inelastic interdot tunnelling in the DQD, mediated by resonant
absorption of an energy quantum from the drive-QPC circuit,
similar to photon assisted tunnelling~\cite{vanderwiel}. The
energy absorbed by the top most DQD electron initially localized
in one dot compensates for the difference of the dots'
electrochemical potentials $\Delta\equiv\mu_L-\mu_R$ and lifts the
Coulomb blockade of interdot and dot-lead tunnelling (see the
insets of fig.~\ref{fig2}c). This picture is further supported by
the observed suppression of $I_\text{DQD}$ inside a small
diamond-shaped region between the resonances (fig.~\ref{fig2}b).
There, the excited state configuration is stable with respect to
dot-lead tunnelling so that absorption of energy doesn't result in
$I_\text{DQD}$~\cite{DQDratchet}.
\begin{figure}[p]
\begin{center}
\includegraphics[clip,width=0.4\linewidth]{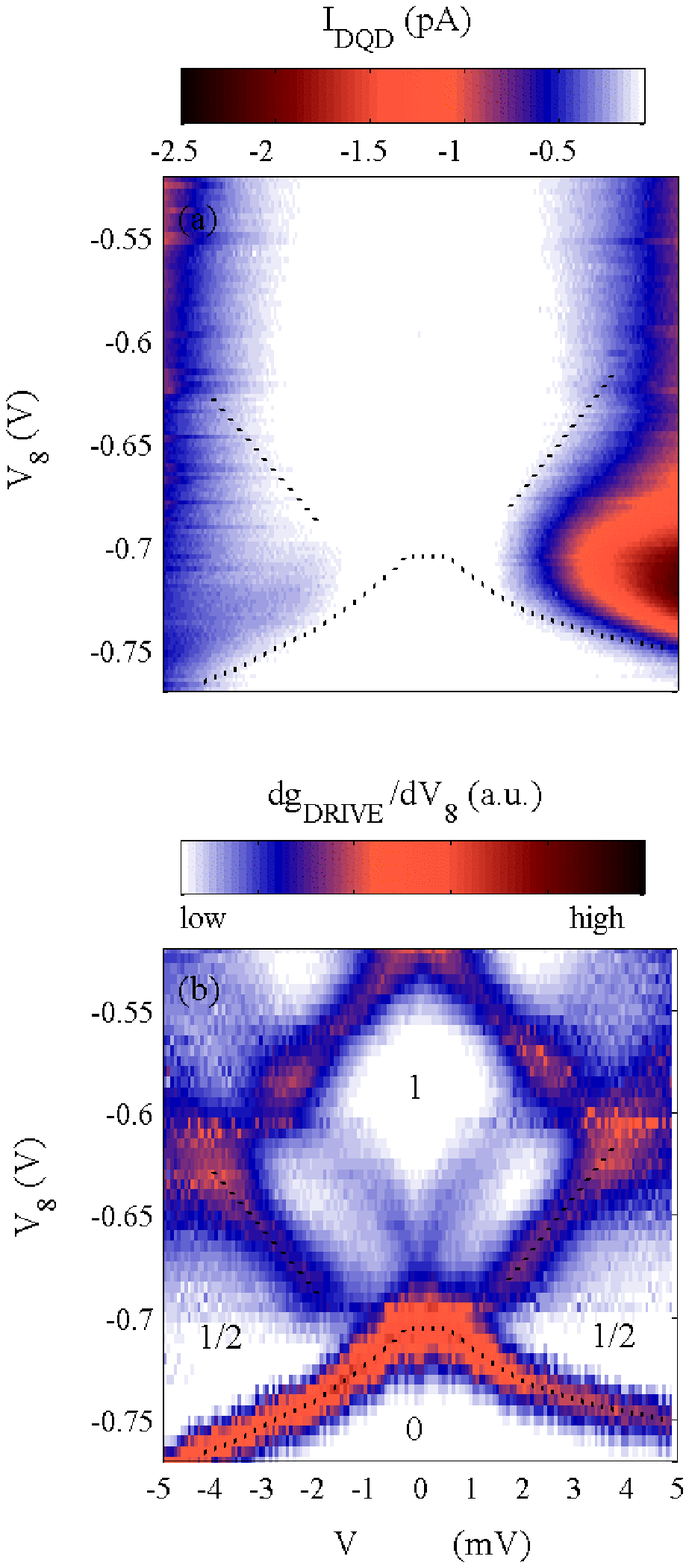} \caption{(a):
Colour plot of $I_\text{DQD}$ vs bias and gate voltage of the
drive-QPC. (b) Colour plot of $g^T_\text{DRIVE}$ for the same
region of the $[V_\text{DRIVE},~V_8]$ plane. Low
$g^T_\text{DRIVE}$ regions are marked by corresponding approximate
values of $g_\text{DRIVE}/G_0$. In both figures dashed guide-lines
mark the boundaries of half integer plateaus on $g_\text{DRIVE}$.}
\label{fig3}
\end{center}
\end{figure}

Owing to the spatial asymmetry of the quantized charge
distribution the DQD  represents a realization of a quantum
ratchet system~\cite{reimann} capable of rectifying nonequilibrium
fluctuations in the environment. The resonant character of the
rectification can be used for spectrometry of the excitation
provided by the drive-QPC. In fig.~\ref{fig2}c we plot
$I_\text{DQD}$ as a function of $\Delta$ along the dashed trace in
the stability diagram of fig.~\ref{fig2}b for a set of
$V_\text{DRIVE}$ values (gate voltage to energy is converted with
a standard calibration procedure~\cite{vanderwiel}). At
$|V_\text{DRIVE}|\gtrsim$~1mV the ratchet contribution to
$I_\text{DQD}$, which is odd in $\Delta$, sets-in within about a
1~meV wide energy band $|\Delta|\lesssim1$~meV.

Obviously, the energy transferred to the detector circuit is a
part of the Joule heat dissipated in the drive circuit. However,
the efficiency of this energization turns out to be a nonmonotonic
function of the drive-QPC conductance. In figure~\ref{fig3}a we
show a colour-scale plot of $I_\text{DQD}$ as a function of
$V_\text{DRIVE}$ and gate voltage $V_8$, which controls the
drive-QPC conductance (fig~\ref{fig1}b). Here,
$\Delta=~-450~\mu$eV. For comparison, a derivative of the
drive-QPC conductance with respect to its gate voltage (below
referred to as transconductance,
$g^T_\text{DRIVE}=dg_\text{DRIVE}/dV_8$) is shown for identical
axes in fig.~\ref{fig3}b. In both figures, the dashed lines mark
the boundaries between the so-called 0.5-plateaus on the
non-linear differential conductance
($g_\text{DRIVE}\approx~G_0/2$) and its pinch-off and first
plateau ($g_\text{DRIVE}\approx~G_0$). The plateaus and the
boundaries between them appear as regions of low and high
transconductance in fig.~\ref{fig3}b~\cite{kristensen}. As follows
from figure~\ref{fig3}a, at fixed $V_\text{DRIVE}$ $I_\text{DQD}$
is maximal on the drive-QPC 0.5-plateau and suppressed on its
first conductance plateau. In other words, the energization of the
DQD ratchet is strong (weak) when the drive-QPC is tuned to a
strongly non-linear (almost linear) transport regime. Note that a
similar, though much less developed, maximum of the energization
efficiency can be observed in the region of the drive-QPC
1.5-plateau at not too high bias~\cite{DQDratchet}.

\section{Counterflow of electrons in isolated QPCs}\label{counterflow}

\begin{figure}[p]
\begin{center}
\includegraphics[clip,width=0.4\linewidth]{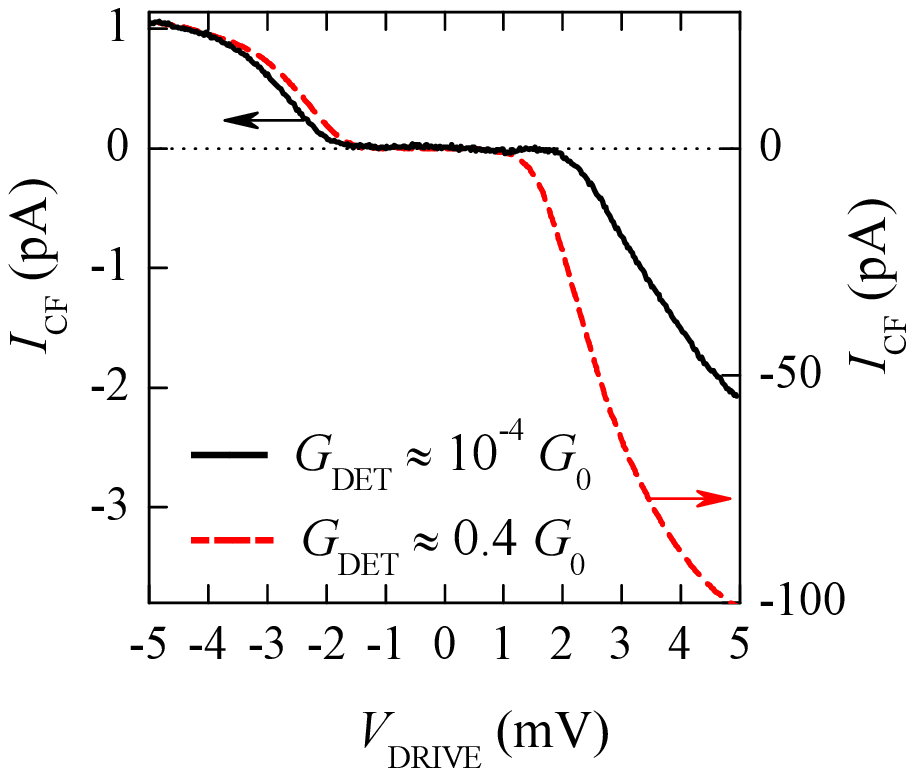} \caption{Current
through the detector-QPC as a function of bias across the
drive-QPC. The data for two indicated values of the detector-QPC
conductance are plotted on corresponding left and right ordinate
scales.} \label{fig4}
\end{center}
\end{figure}
In the previous section we demonstrated that a broad-band energy
transfer from the drive-QPC to the neighbouring circuit can be
detected with a quantum ratchet system. Here we analyse this
energy flow placing a second QPC (detector-QPC) in the detector
circuit, which represents a quantum system with no spatial
asymmetry~\cite{counterflowpaper}. Both drive/detector QPCs have a
one-dimensional (1D) subband spacing of about 4~meV/3~meV, while
the half-width of transition region between the quantized plateaus
is $\delta\approx~0.5$~meV. Throughout this section we keep
$g_\text{DRIVE}\approx0.5~G_0$, which corresponds to the most
pronounced effect. The sketch of the experiment is given in
fig.~\ref{fig1}c. The current generated in the unbiased detector
circuit is measured as a function of $V_\text{DRIVE}$ or gate
voltage $V_3$, which controls the position of the 1D subbands of
the detector-QPC relative to the Fermi energy $E_F$ of its 2DEG
leads (thereby tuning its linear response conductance
$G_\text{DET}$).

\begin{figure}[p]
\begin{center}
\includegraphics[clip,width=0.4\linewidth]{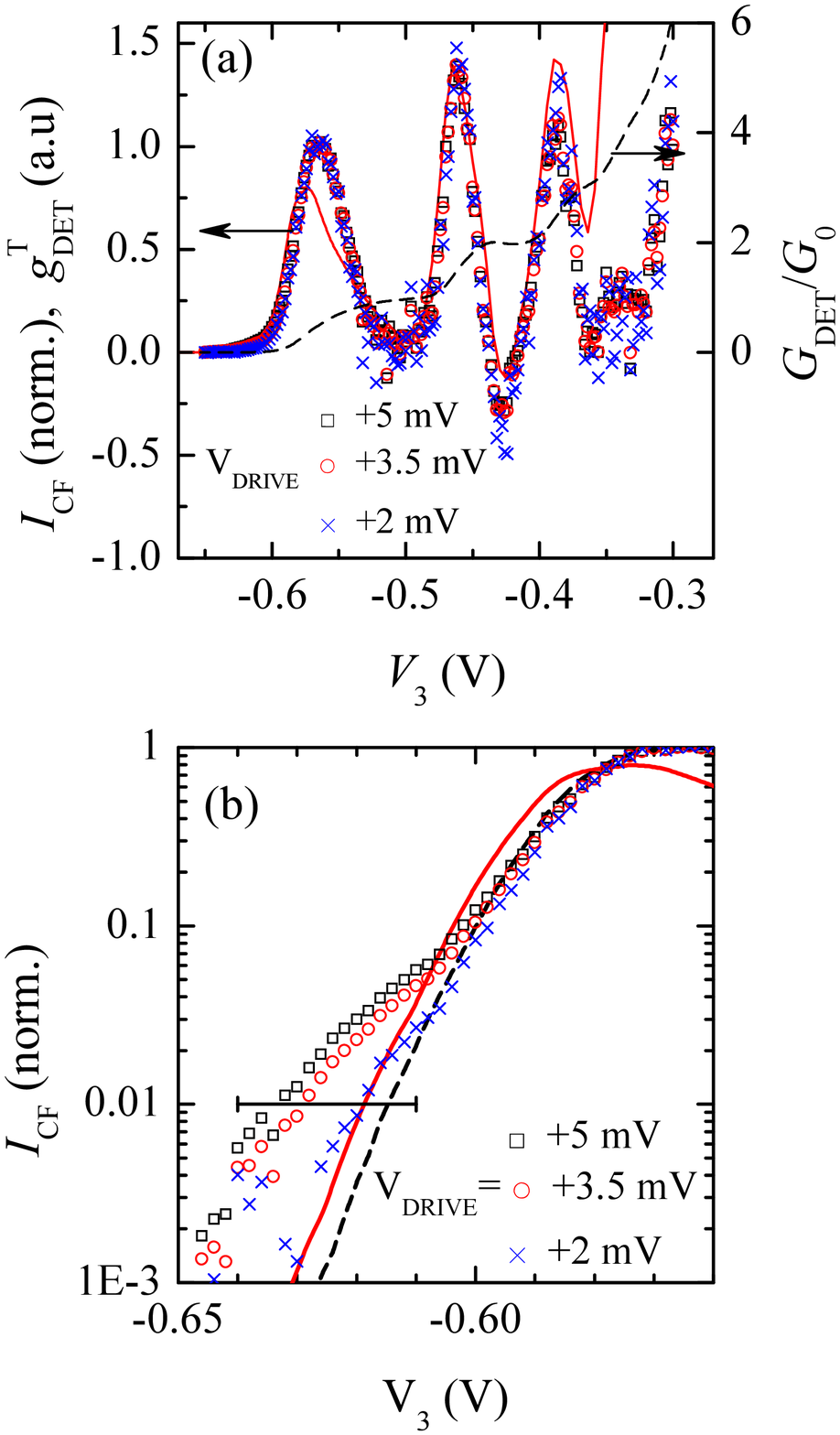} \caption{(a):
$I_\text{CF}$ vs $V_3$ normalized by its leftmost peak value for a
set of $V_\text{DRIVE}$ values (left panel). Conductance (dash)
and transconductance (solid line) of the detector-QPC (right
panel). (b): Log-scale of $I_\text{CF}$ near detector pinch-off.
$g^T_\text{DET}$ (same units as in (a)) and transmission function
$T_0(1-T_0)$ of the detector-QPC  are also shown as solid and
dashed lines respectively. The bar indicates a gate voltage scale
corresponding to a change of $E_0-E_F$ by 1~meV.} \label{fig5}
\end{center}
\end{figure}
The detector current versus $V_\text{DRIVE}$ is plotted in
fig.~\ref{fig4} for two values of $V_3$, which correspond to a
position of the lowest 1D subband bottom $E_0$ well above $E_F$ or
almost aligned with it. At high enough $|V_\text{DRIVE}|$ a finite
current is measured, which is positive/negative for
$V_\text{DRIVE}<0/>0$, i.e. it flows in the direction opposite to
that of $I_\text{DRIVE}$. Below we refer to this current as a
counterflow current $I_\text{CF}$. Note, that $I_\text{CF}$
increases as $E_0$ approaches $E_F$ from above, although much
slower than the relative increment of $G_\text{DET}$. In
figure~\ref{fig5}a we compare the dependencies of $G_\text{DET}$
and $I_\text{CF}$ on $V_3$ in a wide range of gate voltages
between the pinch-off and fully opened detector-QPC. The increase
of $G_\text{DET}$ is accompanied by strong oscillations of
$I_\text{CF}$, which displays three well developed maxima before
the detector-QPC is opened completely. The positions of maxima
correspond to half-integer conductance values
$G_\text{DET}\approx~(i+1/2)G_0$ attained each time the bottom of
the $i$-th 1D subband  $E_i\propto-|e|V_3$ ($i$=0,~1,~2) aligns
with $E_F$.

Oscillations of $I_\text{CF}$ are reminiscent of well-known
oscillations of thermopower in single
QPCs~\cite{molenkamp,dzurak}. In the absence of thermal
equilibrium, the energy balance between the 2DEG leads of the
detector-QPC is broken which results in net electric current:
\begin{equation}
I=\frac{2e}{h}\sum_i\int[f^l_R-f^r_L]T_idE \label{eq1}
\end{equation}
Here, $f^l_R(E)$ ($f^r_L(E)$) is the average occupancy of the left
(right) moving electron states in the right (left) 2DEG lead of
the detector-QPC at energy $E$~\cite{buettiker}. In thermopower
experiments these are just Fermi-Dirac distributions with
appropriate temperatures. The energy dependence of the $i$-th
subband transmission probability evaluated in a saddle-point
approximation~\cite{buettiker} is given by
${T_i=1/(1+\text{exp}([E_i-E_F]/\delta))}$, where $\delta$ is a
half-width of the energy window corresponding to $0.25<T_i<0.75$.
At temperatures low compared to $\delta$ thermoelectric current is
proportional to $\sum dT_i/dE_i$, i.e. it oscillates as the QPC
transconductance. The shape of the oscillations of $I_\text{CF}$
in fig.~\ref{fig5}a is indeed close to that of $g^T_\text{DET}$
(solid line)~\cite{remark0.7}. This indicates that the counterflow
effect is related to energetic imbalance between the two 2DEG
leads of the detector-QPC. Note that a sign change of
$I_\text{CF}$ on the second quantized plateau could be ascribed to
a slightly nonmonotonic behaviour of $G_\text{DET}$ in this region
(i.e. $g^T_\text{DET}<0$).

Despite this qualitative analogy, we find a remarkable
quantitative disagreement between the thermoelectric model and
experiment. In fig.~\ref{fig5}b the counterflow data and
$g^T_\text{DET}$ are plotted on a logarithmic scale near the
pinch-off ($E_0\gg E_F$), where both decay nearly exponentially
with decreasing $V_3$. In this regime, indeed, $g^T_\text{DET}$
can be expressed as
$g^T_\text{DET}\sim\text{exp}(-[E_0-E_F]/\delta)$. Importantly,
$I_\text{CF}$ decays much slower than $g^T_\text{DET}$, which is
readily seen from the data for most negative $V_3$. This means
that electrons excited well above $E_F$ are responsible for the
counterflow in the pinch-off regime, i.e.
$I_\text{CF}\sim\text{exp}(-[E_0-E_F-E^*]/\delta)$, where $E^*$ is
their characteristic excess energy. Standard gate voltage to
energy calibration gives an estimate of $E^*\sim0.5$~meV (see the
scale bar in fig.~\ref{fig5}b corresponding to $\Delta
E_0\approx$1~meV). The thermoelectric model fails to
simultaneously account for the energy scale $E^*$ and measured
$I_\text{CF}$ values~\cite{remarkthermopower}. A leads temperature
difference of about 3~K would be needed in the former case, which
corresponds to thermal currents two orders of magnitude higher
than actually measured (peak values $\sim$10~nA versus
$\sim$100~pA in fig.~\ref{fig4}). Hence, the above analysis shows
that the distribution function of electrons in one of the detector
leads is strongly non-thermal, out-weighted towards high
excitation energies compared to the usual Fermi-Dirac
distribution. The nonequilibrium distribution function is a result
of continuous drive-QPC mediated excitation of a 2DEG region next
to the detector-QPC and its continuous cooling via interchange of
electrons with neighbouring cold 2DEG regions. This process is
accompanied by a non-zero counterflow current across the
detector-QPC thanks to the above mentioned energy dependence of
its transmission probability.

A rough test for a spatial extent of the excited 2DEG region can
be performed by using gates 6 or 10 (instead of the gate 8) to
define the drive-QPC (fig.~\ref{fig1}a). We have checked
~\cite{counterflowpaper} that despite the resulting mutual shift
of the drive and detector QPCs by about $\pm$300~nm along gate C
the counterflow effect is still observed for both directions of
the drive current in each case, confirming the relevance of the
2DEG leads. Finally, the observed direction of the counterflow
defines the following empiric rule. The nonequilibrium lead of the
detector-QPC is the one neighbouring to the drain lead of the
drive-QPC, i.e. the lead with the lower electrochemical potential
where the electrons are being injected (see fig.~\ref{fig1}c).

\section{Excitation of a Quantum Dot with an isolated QPC}\label{QD}

In the last sections we showed how a generation of current occurs
in the DQD- and QPC-based detector circuits neighbouring the
drive-QPC circuit. Here we demonstrate that a nonequilibrium
excitation with a drive-QPC also influences the conductance of a
single QD in the detector circuit. The sketch of the experiment is
shown in fig.~\ref{fig1}d. At fixed $V_\text{DRIVE}$ the
differential QD conductance $g_\text{DOT}$ is measured in the
linear regime as a function of gate voltage $V_2$, which controls
the dot's electrochemical potential. Throughout this section,
again, $g_\text{DRIVE}\approx0.5~G_0$.

\begin{figure}[p]
\begin{center}
\includegraphics[clip,width=0.4\linewidth]{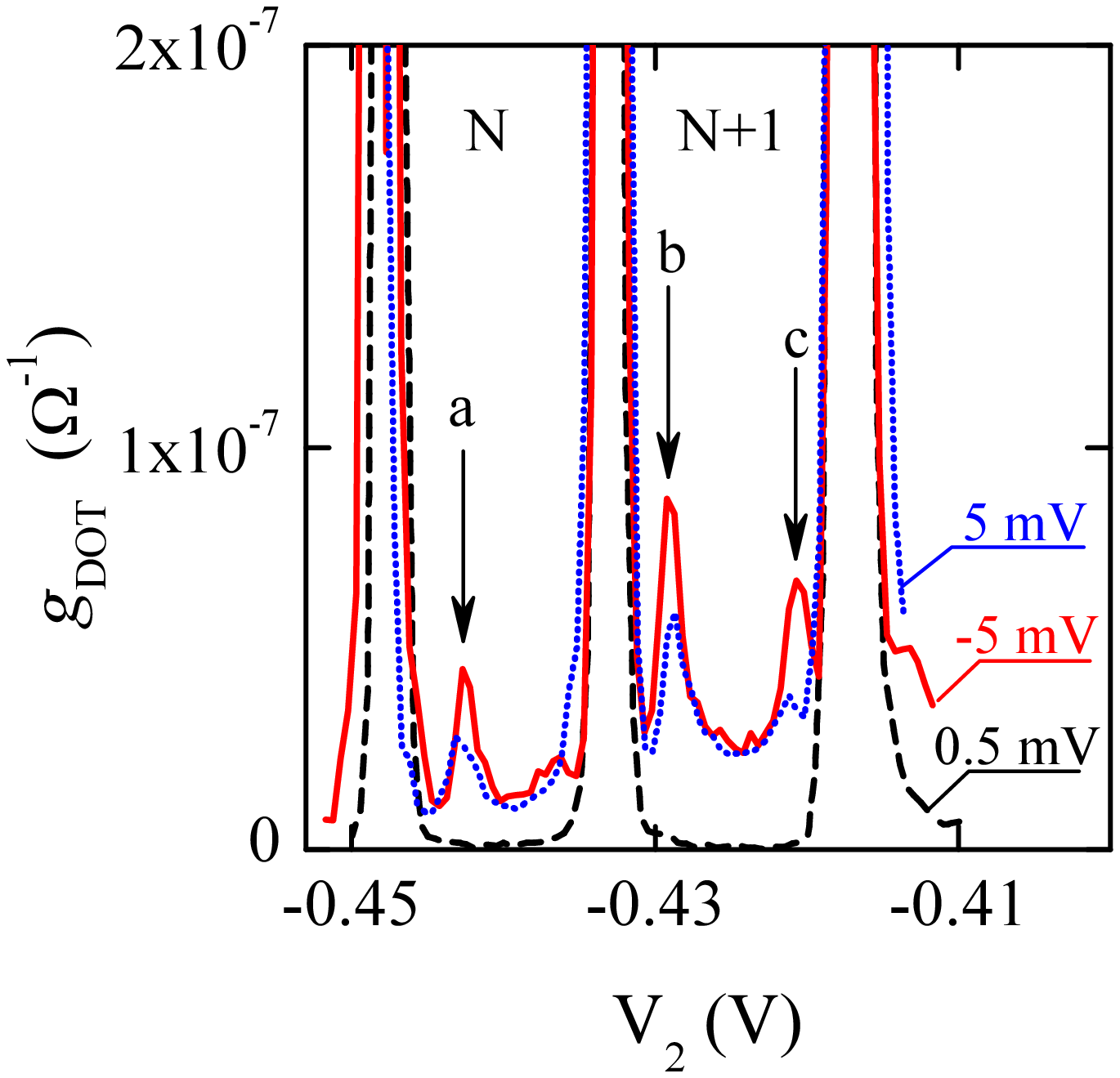}
\caption{Differential conductance of the QD in the detector
circuit vs gate voltage for three values of $V_\text{DRIVE}$.
Three peaks in the Coulomb valleys (arrows a,b,c) correspond to
conductance through excited QD states.} \label{fig6}
\end{center}
\end{figure}

 In fig.~\ref{fig6} $g_\text{DOT}$ is plotted versus
$V_2$ for one relatively small and two much higher values of
$|V_\text{DRIVE}|$. At small drive bias of 0.5~mV nonequilibrium
excitation is ineffective (see two previous sections) and
$g_\text{DOT}$ shows three usual Coulomb blockade peaks. Two
Coulomb valleys between the peaks are marked with numbers N and
N+1 corresponding to the (unknown) total number of QD electrons in
each case. Each of these peaks corresponds to an equilibrium
resonance condition, when the electrochemical potential of the QD
aligns with that of the 2DEG leads ($\mu_\text{LEADS}$). For
instance, for the central peak this condition reads
$E^g_\text{N+1}-E^g_\text{N}=\mu_\text{LEADS}$, where
$E^g_\text{N}$ denotes the total energy  of the ground N-electron
state of the QD. At high $|V_\text{DRIVE}|$ nonequilibrium
excitation lifts the Coulomb blockade and $g_\text{DOT}$ is
strongly increased in Coulomb valleys (at least an order of
magnitude). On top of a smooth background three resonant features
are seen in Coulomb valleys in presence of excitation (see
arrows). These correspond to the transport through the excited
states of the QD.

In presence of excitation the QD is no longer at thermal
equilibrium with its leads and its excited states are occupied
with a probability much higher than that given by usual thermal
fluctuations (exponentially small inside the Coulomb valley). In
this case conductance peaks can be observed at different gate
voltages, compared to the ground state resonances~\cite{Onac}.
E.g. if  $E^*_\text{N}$ denotes the total energy of the excited
N-electron state, an extra conductance peak corresponds to the
resonance condition
$E^*_\text{N}-E^g_\text{N-1}=\mu_\text{LEADS}$. This peak is
shifted to a more positive gate voltage compared to the ground
state Coulomb blockade peak: $\delta
V_2\propto(E^*_\text{N}-E^g_\text{N})$. Similarly, the extra peak
for $E^g_\text{N}-E^*_\text{N-1}=\mu_\text{LEADS}$ is shifted to a
more negative gate voltage: $\delta
V_2\propto-(E^*_\text{N-1}-E^g_\text{N-1})$. The resonances
a/~b/~c in fig.~\ref{fig6} correspond to a set of such
nonequilibrium transitions ${\rm Ex_\text{N}\leftrightarrow
Gr_\text{N-1}/~Ex_\text{N+1}\leftrightarrow
Gr_N/~Ex_\text{N+1}\leftrightarrow Gr_\text{N+2}}$, where $\rm
Ex~(Gr)$ stands for excited (ground) many-electron states. The
excitation energies deduced from the peaks positions equal
$E^*_\text{N}\approx530~\mu$eV and
$E^*_\text{N+1}\approx~340\mu$eV respectively for resonance a and
resonances b,~c.

\begin{figure}[p]
\begin{center}
\includegraphics[clip,width=0.4\linewidth]{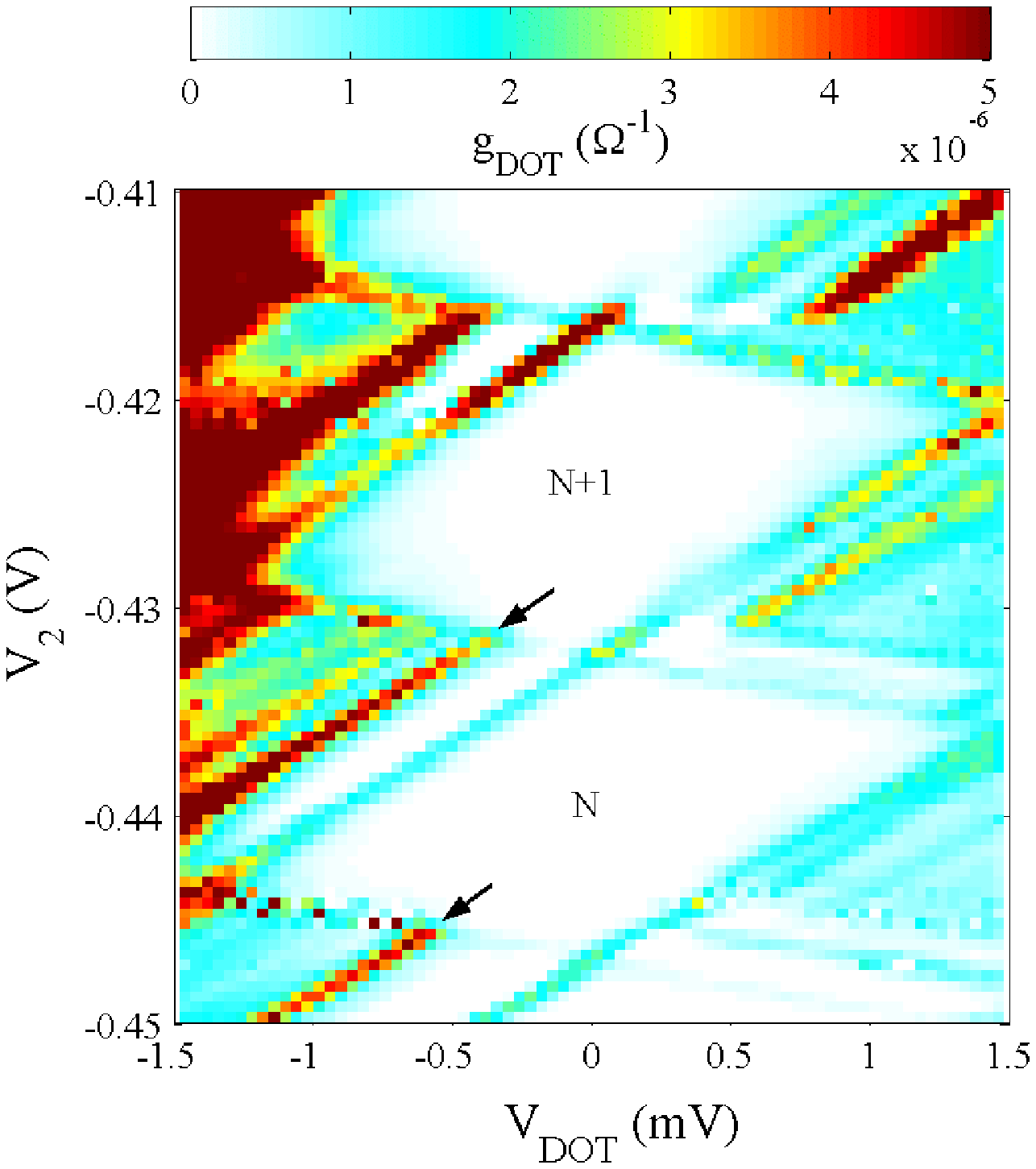}
\caption{Color-scale plot of the QD's differential conductance in
the absence of nonequilibrium excitation with the drive-QPC
($V_\text{DRIVE}=$0.5~mV). Arrows indicate the two strongest
source-resonances corresponding to transport through N+1-electron
and N-electron excited QD states.} \label{fig7}
\end{center}
\end{figure}

Notably, only a few extra resonances are seen in fig.~\ref{fig6}
despite the band-width of the drive-QPC excitation far exceeds a
single-particle level spacing in our QD ($\sim$1~meV versus
$\sim100~\mu$eV). The reason why some resonances are most
pronounced is probably related to optimal (maximal) ratio of the
corresponding dot-lead tunnelling rate to the inelastic relaxation
rate inside the dot. This idea can be directly verified by a
measurement of the non-linear differential conductance of the QD,
where the excited states participate in transport thanks to a
finite bias $V_\text{DOT}$ across the QD. In fig.~\ref{fig7} we
show a color-scale plot of $g_\text{DOT}$ versus $[V_2$,
$V_\text{DOT}]$ in the absence of nonequilibrium excitation.
Diamond shaped regions of Coulomb blockade (Coulomb diamonds) are
marked with corresponding electron numbers (same as in
fig.~\ref{fig6}). X-shaped regions of finite conductance, centered
at the positions of zero bias Coulomb peaks, correspond to gate
voltage range allowed for sequential tunnelling, which grows
proportionally to $|V_\text{DOT}|$. At negative QD bias several
lines of enhanced $g_\text{DOT}$ are distinguished below N+1-th
and N-th Coulomb diamonds, which correspond to different excited
states participating in transport at bias voltages
$|V_\text{DOT}|>E^*-E^g$. The strongest among these resonances
(marked with arrows) indeed correspond to the same excited states
which are responsible for the extra resonances in fig.~\ref{fig6}.

\section{Discussion}\label{conclusions}

In the above sections we demonstrated that the externally biased
drive-QPC can provide a nonequilibrium excitation to the
neighbouring quantum circuit. The excitation has a large bandwidth
of $\sim$1~meV and can be detected with a QD, a DQD or a QPC
placed in the detector circuit. In all three cases a common
feature of the drive-QPC mediated excitation is observed: the
excitation possesses a threshold-like drive-bias dependence and is
suppressed for $V_\text{DRIVE}\lesssim$1~mV. This and other main
observations can be explained in terms of an acoustic-phonon-based
energy transfer mechanism between the two quantum circuits. We
start the discussion from the counterflow effect, which allows a
qualitative argumentation based on the conservation laws.

The key ingredients necessary for the counterflow are
(section~\ref{counterflow}): (i) absorption of energy quanta up to
1~meV by the free 2DEG electrons and (ii) preferential energy flow
to one of the leads. The first requirement allows to rule out a
possible contribution of the photon mediated energy transfer based
on conservation laws. Direct Coulomb interaction between the
electrons of coplanar 2DEGs in the drive and detector circuits
also cannot account for the counterflow. The momentum transferred
via Coulomb interaction is restricted by the minimum distance
between the electrons $|q|\leq10^{-5}\text{cm}^{-1}$ (gate C wider
than 100~nm, see fig.~\ref{fig1}a), which is much smaller than the
Fermi momentum $k_F>10^{-6}\text{cm}^{-1}$. Under such conditions
only a forward Coulomb scattering can occur, i.e. a mutual
scattering of the two electrons moving in the same direction in
different circuits, which obviously cannot give rise to the
counterflow.

Both above conditions can be satisfied if electron-phonon
interaction is taken into account. Thanks to its ballistic nature
the current flowing across the drive-QPC results in injection of
hot electrons above the Fermi energy into the drain lead, which
leave unoccupied states (holes) below the Fermi energy in the
source lead~\cite{remarksourcelead}. The drain electron and source
hole excess energies (referred to respective Fermi energies
$\epsilon^D_e=E-E^D_F>0$ and $\epsilon^S_h=E-E^S_F<0$) satisfy
$\epsilon_e+|\epsilon_h|=|eV_\text{DRIVE}|$. Energy relaxation of
nonequilibrium carriers in the drive-circuit can occur via
emission of acoustic phonons. Part of the phonons with momenta
parallel to the interface can be re-absorbed in the nearby
detector circuit. Such phonons have momenta up to $2\hbar k_F$ and
energies up to $2\hbar k_Fv_s\approx$~0.6~meV (calculated for
sound velocity $v_s=3\cdot10^5\text{cm/s}$) and can give rise to a
strongly nonequilibrium distribution of electrons in the detector.
Particularly important for the counterflow effect is a non-linear
transport regime across the drive-QPC near its
pinch-off~\cite{counterflowpaper}. Here the excess energies of the
injected drain electrons are much higher than those of the source
holes $\epsilon^S_h\ll\epsilon^D_e\approx|eV_\text{DRIVE}|$, so
that the emission of phonons in the drive circuit occurs
preferably at the drain side~\cite{heiblum}. Because of the device
geometry (fig.~\ref{fig1}a), absorption of phonons in this case
happens preferably in the neighbouring lead of the detector
circuit. This naturally explains the origin of the asymmetric
excitation responsible for the counterflow and the sign of this
effect. Additional support to the above discussed mechanism comes
from the near independence of the effect on the physical distance
between the drive and detector QPC, which was controlled by the
voltage applied to the gate C~\cite{ep2dsprocs} (see the sketch of
fig.~\ref{fig1}c).

Next we speculate how the acoustic-phonon-based energy transfer
mechanism could explain our observations for the DQD quantum
ratchet and QD excitation. In principle, high energy acoustic
phonons can be directly absorbed by the localized QD
electrons~\cite{fedichkin}, which would suffice for a qualitative
explanation. However, there exists an alternative microscopic
mechanism. Strongly nonequilibrium electrons in the detector
circuit create high frequency electric field fluctuations, which
can in turn drive inelastic transitions in a QD and a DQD. In
fact, the data of fig.~\ref{fig2}c and fig.~\ref{fig6} look very
similar to photon-assisted tunnelling data in DQD and QD under
microwave excitation~\cite{vanderwiel,PAT}. An important hint in
favor of the latter mechanism is the observation of the
$\Delta$-independent and counterflow-like contribution to the
drive-QPC mediated current through the DQD~\cite{DQDratchet}.
Still, it is hard to unambiguously determine which of the two
microscopic mechanisms is more relevant for the excitation of the
DQD ratchet and the QD in our experiments.

While the spatial asymmetry of the excitation in the drive
circuit, characteristic for the non-linear transport regime,  is
relevant for the counterflow, it is not necessary for the QD and
DQD ratchet experiments.  Therefore one would naively expect the
phonon-mediated excitation to be efficient also at small drive
bias in the last two experiments. In contrast, we find that in all
three cases the drive-QPC mediated excitation is suppressed for
$|V_\text{DRIVE}|\lesssim1$~mV~\cite{suppression} (see, e.g.,
fig.~\ref{fig2}). Though it is hard to give a quantitative
explanation for this onset, we simply attribute it to the
steepness of the drive-bias dependence owing to a rapid decrease
of an electron-phonon energy relaxation rate at small excess
energies. In the so-called Bloch-Gr\"{u}neisen limit a cooling
power of the 2DEG can fall as $P\sim T_e^3-T_l^3$ or faster at low
temperatures in a polar crystal like GaAs~\cite{ridley}
($T_e,~T_l$ are the electron and lattice temperatures). Hence a
cooling power of the drive-circuit falls at not too high bias as
$P\sim \alpha^3|V_\text{DRIVE}|^3$, where $\alpha\leq1$ is a bias
lever-arm coefficient, which defines the characteristic excess
energy of the nonequilibrium carriers ($\alpha=1$ in the strongly
non-linear regime, see above). The average path length a
nonequilibrium electron travels before emitting an acoustic phonon
at small excess energies can exceed even the size of our whole
device~\cite{ridley}. This should result in even steeper drive
bias dependence of the detector response, since a vanishingly
small fraction of the phonons emitted in the drive-circuit can be
re-absorbed in the vicinity of the neighbouring detector
nanostructure as $|V_\text{DRIVE}|$ is decreased. In the end, we
would like to point out that the above qualitative argument alone
fails to fully explain some our observations, e.g., the enhanced
efficiency of the DQD ratchet excitation near the drive-QPC
pinch-off (fig.~\ref{fig3}a). Possibly some properties of the
drive-QPC in the non-linear transport regime and/or an alternative
mechanism of the energy transfer could be relevant here, see
e.g.~\cite{chudnovskiy}.

In conclusion, we studied the energy transfer from an externally
biased drive circuit containing a drive-QPC to a neighbouring
detector circuit containing a DQD, a QPC or a QD. In all three
cases a 1~meV bandwidth excitation is observed, provided the drive
bias is in the mV range. The main features of the experiments are
explained within a qualitative model of acoustic-phonon-based
energy transfer mechanism. Non-equilibrium acoustic phonons are
emitted in the vicinity of the drive-QPC and re-absorbed in the
2DEG of the detector circuit. This mechanism is most efficient at
high drive bias and near the drive-QPC pinch-off, which has to be
considered in experiments on coupled quantum circuits.

\section*{Acknowledgements}

The authors are grateful to V.T.~Dolgopolov, A.W.~Holleitner,
C.~Strunk, F.~Wilhelm, I.~Favero, A.V.~Khaetskii,
N.M.~Chtchelkatchev, A.A.~Shashkin, D.V.~Shovkun and P.~H\"anggi
for valuable discussions and to D.~Schr$\ddot{\text{o}}$er and
M.~Kroner for technical help. We thank the DFG via SFB 631, the
BMBF via DIP-H.2.1, the Nanosystems Initiative Munich (NIM) and
VSK the A.~von~Humboldt foundation, RFBR, RSSF, RAS  and the
program "The State Support of Leading Scientific Schools" for
support.


\begin{thebibliography}{99}
\bibitem{remark0.7} A poor comparison between $I_\text{CF}$ and
$g^T_\text{DET}$ for the left-most peak in fig.~\ref{fig5}a is
related to the so-called 0.7 anomaly in detector conductance,
which causes the asymmetry of the $g^T_\text{DET}$ peak. This
effect is usually attributed to interactions, hence the
single-particle approximation to thermopower is not likely to be
accurate in this region.
\bibitem{remarkthermopower} We note that the saddle-point
model~\cite{buettiker} used is not fully consistent with the
experimental dependence of $G_\text{DET}$ on $V_3$. E.g. a
proportionality $g^T_\text{DET}\propto T_0\cdot(1-T_0)$ predicted
by the model is not exactly fulfilled in our experiment (compare
solid and dashed lines in fig.~\ref{fig5}b). Such minor
discrepancies do not effect our conclusions about the relation
between thermopower and counterflow.
\bibitem{remarksourcelead} Here we choose the source/drain to be the
lead with the higher/lower electrochemical potential, so that
$\mu^S_\text{DRIVE}=\mu^D_\text{DRIVE}+|eV_\text{DRIVE}|$ (note
the definition dependence on the sign of $V_\text{DRIVE}$).
\bibitem{suppression} Note, that suppression of the excitation at low bias
doesn't allow us to observe a minimum bias condition for
single-quanta excitation at certain energy
$|V^\text{min}_\text{DRIVE}|\geq|E/e|$, which was verified in
shot-noise detection~\cite{gustavsson,Onac}.
\end{thebibliography}
\end{document}